\documentclass[a4paper,fleqn,10pt]{article}
\usepackage[numbers,sort&compress]{natbib}
\usepackage{graphicx}
\usepackage{subfigure}
\usepackage{latexsym}
\usepackage{amsmath,amssymb}
\usepackage{dcolumn}
\usepackage{float}
\usepackage{caption}

\begin{document}
\captionsetup[figure]{labelfont={bf},labelformat={default},labelsep=space,name={Fig.}}

\title{{\bf Quasinormal modes, Hawking radiation and absorption of the massless scalar field for Bardeen black hole surrounded by perfect fluid dark matter}
\author{\normalsize Qi Sun$^{1}$, Qian Li$^{1}$, Yu Zhang$^{1}$\thanks{Corresponding Author(Y. Zhang): Email: zhangyu\_128@126.com}, Qi-Quan Li$^{1}$\\
   \normalsize \emph{$^{1}${Faculty of Science, Kunming University of Science and Technology, }}\\ \normalsize \emph {Kunming, Yunnan 650500, People's Republic of China}}}

\date{}
\maketitle \baselineskip 0.3in

{\bf Abstract} \ {We study the quasinormal modes, Hawking radiation and absorption cross-section of the Bardeen black hole surrounded by perfect fluid dark matter for a massless scalar field.
		Our results show that the oscillation frequency of quasinormal modes is enhanced as magnetic charge $g$ or the dark matter parameter $\alpha$ increases. For damping rate of quasinormal modes, the influence of them is different. Specifically, the increase of dark matter parameter $\alpha$ makes the damping rate increasing at first and then decreasing. While the damping rate is continuously decreasing with the increase of the magnetic charge $g$. Moreover, we find that the increase of the dark matter parameter $\alpha$ enhances the power emission spectrum whereas magnetic charge $g$ suppresses it. This means that the lifespan of black holes increases for smaller value of $\alpha$ and larger value of $g$ when other parameters are fixed. Finally, the absorption cross-section of the considered black hole is calculated with the help of the partial wave approach.
		Our results suggest that the absorption cross-section decreases with the dark matter parameter $\alpha$ or the magnetic charge $g$ increasing.}

{\bf Key words:} \ {Black hole; quasinormal modes; Hawking radiation; absorption cross-section; massless scalar field.}


\section{Introduction}\label{sec-introduction}

In the past few years, LIGO-Virgo detected gravitational waves from a binary black hole merger \cite{LIGOScientific:2016aoc,LIGOScientific:2016sjg,LIGOScientific:2018mvr} and Event Horizon Telescope captured the shadow image of supermassive black hole in the center of M87*, which is a giant elliptical galaxy \cite{EventHorizonTelescope:2019dse,EventHorizonTelescope:2019ths}. These observations have excellent concordance with the prediction of general relativity concerning the existence of black hole. But the existence of singularity in black hole is an attractive problem in general relativity. At the singularity of spacetime, the curvature and density tend to infinity and the laws of physics are invalid. In other words, the presence of singularity denotes that the predictability of physical laws break down near the singularity, which marks the incompleteness of general relativity.
Fortunately, regular black holes (nonsingular) are proposed to avoid the singularities by some attempts, for example, the theory of quantum gravity replaces the classical theory and matter backgrounds take the place of central singularity. Inspired by some ideas, Bardeen firstly constructed a regular black hole \cite{Bardeen}, called Bardeen black hole. Ayon-Beato \textit{et al.} pointed out that the source of Bardeen black hole is a nonlinear electrodynamic field \cite{Ayon-Beato:2000mjt}. Since the first regular black hole model was proposed, the study of regular black holes has never stopped. There are more regular black holes that have been derived such as Dymnikova black hole \cite{Dymnikova:1992ux}, Bronnilov black hole \cite{Bronnikov:2000vy}, Hayward black hole \cite{Hayward:2005gi} and the generic regular nonlinear electrodynamics black holes \cite{Fan:2016hvf}. Moreover, Toshmatov and his collaborators generalized the solutions of regular black holes \cite{Fan:2016hvf} to the case of rotating based partly on Newman-Jains algorithm.\cite{Toshmatov:2017zpr} The proposal of these regular black holes has stimulated further research on the properties of these black holes, for example, the shadow of black hole \cite{Tsukamoto:2017fxq,Kumar:2019pjp,Ahmed:2020jic} and black hole thermodynamics \cite{Saadat:2013rna,Breton:2015iev,Jawad:2020hju}.

	When Oort used a telescope to observe stars in the Milky Way, he found that they were moving so fast that they could escape the gravitational effects of all the visible matter in the galaxy \cite{OortJH}.
	It denotes that our universe  should need more matter except  the visible matter for the sake  of explaining this astronomical  observation. Besides, the researchers could not directly perceive the existence of the matter which provides ``more mass'' in our universe by ordinary means. So they assumed that the matter does not emits light and then called it dark matter. Then Rubin \textit{et al.} found that stars far from the center of the galaxy spin significantly faster than expected \cite{Rubin:1970zza} and the image of Bullet Cluster was observed by Observatories in 2006 \cite{Clowe:2006eq}. The observations both support the hypothesis of dark matter. Up to now, it is generally accepted that our current universe consists mainly of 4.9$\%$ baryon matter, 26.8$\%$ dark matter, and 68.3$\%$ dark energy \cite{Planck:2015fie}. And scholars further defined dark matter being non-baryonic as well as non-luminous. The properties of black holes surrounded by dark matter have been studied including the thermodynamics \cite{Jamil:2009eb}, weak deflection angle \cite{Pantig:2022toh,Pantig:2021zqe}, shadow \cite{Pantig:2021zqe,Konoplya:2019sns} and geodesic motion. \cite{Stuchlik:2022xtq}
	As mentioned above, it is thus clear that dark matter has always been a hot topic. Therefore, it is necessary to study black holes under dark matter. Perfect fluid dark matter (PFDM) as a candidate of dark matter was proposed by Kiselev \cite{Kiselev:2002dx,Kiselev:2003ah}. Moreover, he constructed a new class of spherical asymmetric solutions surrounded by PFDM \cite{Kiselev:2003ah}, which explains the rotation curves in spiral galaxies by asymptotic behavior of quintessential matter \cite{Kiselev:2002dx,Kiselev:2003ah,Toshmatov:2015npp,Chakrabarty:2018skk} at larger distances. Then Heydarzade and Darabi derived charged/uncharged Kiselev-like black holes in Rastall theory \cite{Heydarzade:2017wxu}. Xu \textit{et al.} obtained Kerr-Newman-anti-de Sitter black hole surrounded by PFDM in Rastall gravity \cite{Xu:2017vse}. In this paper, we take one of regular black holes immersed in dark matter into consideration, i.e., Bardeen black hole surrounded by PFDM proposed by Zhang and his cooperators \cite{Zhang:2020mxi}.

	Now, the model of Bardeen black hole surrounded by PFDM has been built. But how should we know the properties of the black hole? A black hole is not an isolated object and it interacts with its environment. When binary black hole merger, it can produce gravitational waves. And the signal of it can be divided into three phases: inspiral, merger and ringdown.
	At ringdown phase, the black hole emits gravitational waves having the form of discrete and complex frequencies, which are called quasinormal modes (QNMs) \cite{Berti:2005ys}. As the ``characteristic sound" of black holes, the real part of quasinormal frequencies indicates the oscillation frequency while the imaginary part of it represents damping rate. It is noted that QNMs only rely on black hole parameters and the type of perturbation field, which are classical ``fingerprints" of black holes. Therefore, we can obtain the information of black holes by solving the wave equation under special boundary, i.e., QNMs. Vishveshwara first pointed out that the signal decays with the curve of exponential for most of the time, which comes from perturbed black hole \cite{Vishveshwara:1970zz}. Then Chandrasekhar and Detweiler discussed in detail the QNMs of Schwarzschild black hole \cite{Chandrasekhar:1975zza}. After that, a number of researches have sprung up on QNMs \cite{Setare:2003hm,Setare:2003bd,Setare:2004rt,Setare:2004uu,Zhang:2006ij,Stefanov:2010xz,Stefanov:2010qwy,Toshmatov:2015wga,Konoplya:2020bxa,Ma:2022gzr,Konoplya:2020jgt,Pantig:2022gih}, which are aimed to analyze the parameters of black hole, detect the connection of QNMs with area spectrum and the relation of QNMs with gravitational lens.

	On the other hand, Hawking proposed black holes can emit thermal radiation when quantum effects are considered and this radiation is called Hawking radiation. As quantum ``fingerprints" of black holes, Hawking radiation has the information about the evaporation dynamics of black holes. And the process of Hawking radiation is that negative particles enter black holes and positive particles escape to infinity, which is produced by a vacuum fluctuation near the event horizon of black holes \cite{Hawking:1975vcx}. It should be pointed that the emergence of Hawking radiation is a combination of general relativity, quantum mechanics and thermodynamics. Since the concept of Hawking radiation was proposed, the researches on the topic have not stopped in the last few decades. In 1975, Damour and Ruffini developed a method for computing Hawking radiation \cite{Damour:1976jd}. In 2000, Parikh and Wilczek proposed a quantum tunneling method to research Hawking radiation \cite{Parikh:1999mf} and then Kerner and Mann further improved the method \cite{Kerner:2008qv,Kerner:2007rr}.
	In 2011, Kokkotas \textit{et al.} used 6th-order Wentzel-Kramers-Brillouin (WKB) method proposed by Konoplya to compute Hawking radiation by gray-body factors \cite{Konoplya:2003ii,Kokkotas:2010zd}. Moreover, due to the feature of versatility and flexibility, the 6th order WKB method is adopted to obtain Hawking radiation in many papers \cite{Konoplya:2020jgt,Konoplya:2020cbv,Li:2022wzi} and we also employ the method in this paper.

	The forms of general relativity and quantum mechanics are incompatible, which are theories of gravitational interaction and atomic scale, respectively. However, since Hawking discovered that black holes can absorb particles, the topic of absorption cross-section of black hole for different test fields has spurred interest of scholars \cite{Li:2022jda,Ma:2022gzr,Okyay:2021nnh,Pantig:2022ely}. After obtaining the analytical solution of the wave equation, Sanchez discovered that the absorption cross-section for a massless scalar field oscillates in the vicinity of geometric-optics limit \cite{Sanchez:1977si}. Unruh investigated the absorption cross-section of nonrotating black hole. And he found the absorption cross-section for Dirac particles is 1/8 of massive scalar particles in the low-energy situation \cite{Unruh:1976fm}. In the Schwarzschild spacetime, Crispino \textit{et al.} numerically calculated the absorption of electromagnetic waves for arbitrary frequencies \cite{Crispino:2007qw}. Further, Macedo and Crispino studied the absorption of Bardeen black hole and compared it with Reissner-Nordstr\"{o}m black hole for a massless scalar field. Then they found the absorption cross-section of the two black holes might be the same at high frequency \cite{Macedo:2014uga}.

	The remainder of the paper is organized as follows. In Sec.~\ref{sec-static}, we review the Bardeen black hole surrounded by PFDM and discuss the effective potential with respect to different parameters. In Sec.~\ref{sec-massless}, the 6th-order WKB method is briefly introduced and used to calculate QNMs of a massless scalar field for the black hole under consideration. In Sec.~\ref {sec-hawking}, we consider the Hawking temperature and calculate the energy emission rate in the black hole.  In Sec.~\ref{sec-absorption}, using the partial wave method, we investigate the absorption cross-section of the black hole. Conclusions are presented in the last section. Moreover, we set $M$=$G$=$c$=1 in this paper.

\section{Massless scalar field equation in static and spherically symmetric Bardeen black hole surrounded by PFDM}\label{sec-static}

For reasonably explaining astronomical observations at different scales, such as the Cosmic Microwave Background and galaxy rotation curve, the concept of dark matter has been proposed. Kiselev \cite{Kiselev:2002dx,Kiselev:2003ah} proposed PFDM as one of alternative models of dark matter, which has the characters of perfect fluid, i.e.,  isotropic pressure and mass density. Assume that the black hole is immersed in PFDM, Zhang \textit{et al.} \cite{Zhang:2020mxi} considered the coupling of the gravitational and a nonlinear electromagnetic field, and obtained the solution of Bardeen balck hole surrounded by PFDM. Now let us briefly review their work. After the mentioned above two fields are coupled, the Einstein-Maxwell equations can be described as follows:
	\begin{gather}
		G_{\mu}^{\,\nu}=2\left(\frac{\partial\mathcal{L}\left(F\right)}{\partial F}F_{\mu\lambda}F^{\nu\lambda}-\delta_{\mu}^{\,\nu}\mathcal{L}\right)+8\pi T_{\mu}^{\,\nu},\label{eq:EM1}\\
		\nabla_{\mu}\left(\frac{\partial\mathcal{L}\left(F\right)}{\partial F}F^{\nu\mu}\right)=0,\label{eq:EM2}\\
		\nabla_{\mu}\left(*F^{\nu\mu}\right)=0,\label{eq:EM3}
	\end{gather}
	where $F_{\mu\nu}=2\nabla_{[\mu}A_{\nu]}$ and $\mathcal{L}$ is a function of $F\equiv\frac{1}{4}F_{\mu\nu}F^{\mu\nu}$ defined by \cite{Ayon-Beato:2000mjt}
	\begin{equation}
		\mathcal{L}\left(F\right)=\frac{3M}{\vert g\vert^{3}}\left(\frac{\sqrt{2g^{2}F}}{1+\sqrt{2g^{2}F}}\right)^{\frac{5}{2}},
	\end{equation}
	in which $g$ denotes magnetic charge, and $M$ indicates the black hole mass. The energy-momentum tensor can be expressed as $T_{\nu}^{\,\mu}=\mathrm{diag}(-\epsilon,p_r,p_\theta,p_\phi)$ for a black hole surrounded by PFDM, with the density, radial and tangential pressures of the dark matter being \cite{Kiselev:2002dx,Li:2012zx}
	\begin{equation}
		-\epsilon=p_r=\frac{\alpha}{8\pi r^3}\qquad \text{and} \qquad p_\theta=p_\phi=-\frac{\alpha}{16\pi r^3}.
	\end{equation}

	Assume that the metric is static and spherically symmetric, which can be represented in the following form:
	\begin{equation}
		\label{eq:spherically metric}
		{ds}^{2}=-f(r) {d} t^{2}+f(r)^{-1} {~d} r^{2}+r^{2} {~d} \Omega^{2},
	\end{equation}
	with
	\begin{equation}\label{eq:f}
		f\left(r\right)=1-\frac{2Mr^{2}}{\left(r^{2}+g^{2}\right)^{\frac{3}{2}}}+\frac{\alpha}{r}\,\ln{\frac{r}{\vert\alpha\vert}},
	\end{equation}
	where $\alpha$  denotes the parameter for the density and pressure of PFDM, called the dark matter parameter. And Zhang \textit{et al.} have shown that the weak energy condition is satisfied when $\alpha<0$ \cite{Zhang:2020mxi}. From above equation, it can be seen that the black hole recoveres to Schwarzschild black hole when dark matter parameter $\alpha=0$ and magnetic charge $g=0$. When dark matter parameter $\alpha$ vanishs, it reduces to Bardeen black hole. And it restores to Schwarzschild black hole immered by PFDM when $g=0$.

	The general covariant equation of a massless scalar field in curved spacetime can be given by
	\begin{equation}
		\frac{1}{\sqrt{-g}}\partial_\mu \left(\sqrt{-g}g^{\mu \nu}\partial_\nu\Phi\right)=0.
		\label{KGg}
	\end{equation}
	
	For separating the radial and angular directions, we introduce
	\begin{equation}\label{tr}
		\Phi(t,r,\theta,\phi)=\frac{1}{r}\,e^{-i\omega t}Y_{l}(\theta,\phi)\Psi(r),
	\end{equation}
	and
	\begin{equation}\label{rx}
		dr_*=\frac{dr}{f(r)}.
	\end{equation}
	
	Here $Y_{l}(\theta,\phi)$ and $r_*$ are the spherical harmonics function and tortoise coordinate, respectively. And then we substitute Eqs.~(\ref{eq:spherically metric}), (\ref{tr}) and (\ref{rx}) into Eq.~(\ref{KGg}), so a Schr\"{o}dinger-like equation representing a massless scalar field perturbation in the considered black hole spacetime can be expressed as
	\begin{equation}\label{wave-equation}
		\frac{d^2\Psi}{dr_*^2}+\left(\omega^2-V(r)\right)\Psi=0,
	\end{equation}
	where
\begin{equation}\label{scalarpotential}
\begin{aligned}
V(r)=& \left(1-\frac{2 M r^{2}}{\left(g^{2}+r^{2}\right)^{3 / 2}}+\frac{\alpha \ln \left(\frac{r}{\vert\alpha\vert}\right)}{r}\right) \\
& \times \left(\frac{l(l+1)}{r^{2}}+\frac{6 M r^{2}}{\left(g^{2}+r^{2}\right)^{5 / 2}}-\frac{4 M }{\left(g^{2}+r^{2}\right)^{3 / 2}}+\frac{\alpha}{r^{3}}-\frac{\alpha \ln \left(\frac{r}{\operatorname{\vert\alpha\vert}}\right)}{r^{3}}\right).
\end{aligned}
\end{equation}

\begin{figure}
		\centering
		\includegraphics[width=0.45\linewidth]{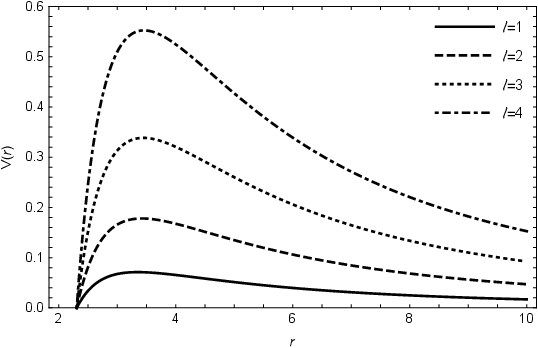}
		\caption{The behavior of effective potential versus $r$ for $l=1, 2, 3, 4$ with fixed  $g=0.1$, $\alpha=-0.1$.}
		\label{vl}
	\end{figure}

\begin{figure}
		\begin{minipage}[t]{0.45\linewidth}
			\includegraphics[width=\linewidth]{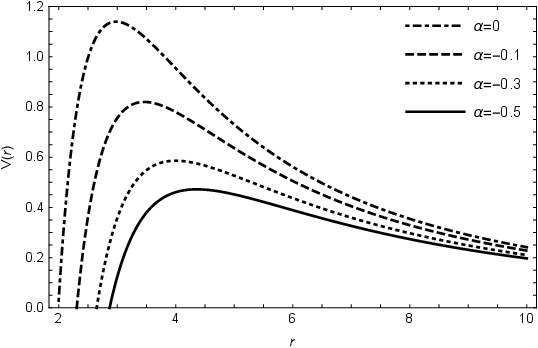}
			\caption{The behavior of effective potential versus $r$ for $\alpha=-0.5, -0.3, -0.1, 0$ (Bardeen black hole) with fixed $g=0.1$, $l=5$.}
			\label{va}
		\end{minipage}%
		\hfill%
		\begin{minipage}[t]{0.45\linewidth}
			\includegraphics[width=\linewidth]{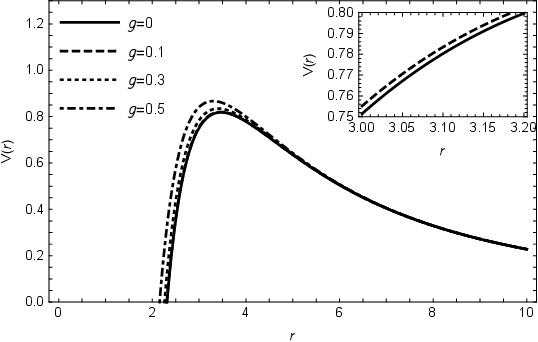}
			\caption{The behavior of effective potential versus $r$ for $g=0$ (Schwarzschild black hole surrounded by PFDM), 0.1, 0.3, 0.5 with fixed  $\alpha=-0.1$, $l=5$.}
			\label{vg}
		\end{minipage}
	\end{figure}

	We can see that the effective potential $V(r)$ depends on multiple number $l$, dark matter parameter $\alpha$ and magnetic charge $g$. In Figs. \ref{vl}-\ref{vg}, we show the variation of the effective potential with $r$ under different parameters. At the same value of the parameters $\alpha$ and $g$, the peak value of potential barrier increases and the peak location moves to right with the increase of $l$.
	For fixed $g$ and $l$, the barrier height of the effective potential decreases obviously with the increase of the absolute value of $\alpha$, and the peak of the effective potential shifts to the right.
	The peak value of the effective potential increases as $g$ increases, and the location of peak moves to the left for fixed $\alpha=-0.1$ and $l=5$. Compared  Fig. \ref{va} and Fig. \ref{vg}, we can find that dark matter parameter has greater influence than magnetic charge for the effective potential.

\section {QNMs of a massless scalar field perturbation}\label{sec-massless}
	
	In this section, we plan to study the relationship of QNMs with overtone number $n$, spherical harmonic index $l$, dark matter parameter $\alpha$ and magnetic charge $g$, respectively. In order to calculate the QNMs of Bardeen black hole surrounded by PFDM for a massless scalar field, we adopt the WKB method. Schutz and Will who considered that the perturbation equation for a particle has the form of Schr\"{o}dinger-like equation, proposed a semi-analytical method to obtain the QNMs, i.e., WKB method and found its accuracy is about 6$\%$ \cite{Schutz:1985km}. Iyer and Will expanded it to the 3rd-order letting its accuracy up to fractions of 1$\%$ for the fundamental mode \cite{Iyer:1986np}.
	Then Konoplya promoted it to the 6th-order \cite{Konoplya:2003ii} and the 13th-order is developed by Matyjasek and Opala \cite{Matyjasek:2017psv}. But it should be noted that WKB method has a characteristic of asymptotic convergence, its accuracy cannot be improved simply by increasing the order of formula. To pursue the efficiency and accuracy of calculation, we exploit the 6th-order WKB method to compute the QNMs of the black hole, which can be written as \cite{Konoplya:2003ii}
	\begin{equation}
		\label{w}
		i \frac{\omega_{n}^{2}-V_{0}}{\sqrt{-2 V_{0}^{\prime \prime}}}-\sum_{i=2}^{6} \Lambda_{i}=n+\frac{1}{2}.
	\end{equation}
	
\begin{figure*}
		\begin{center}
			\includegraphics[width=0.45\textwidth]{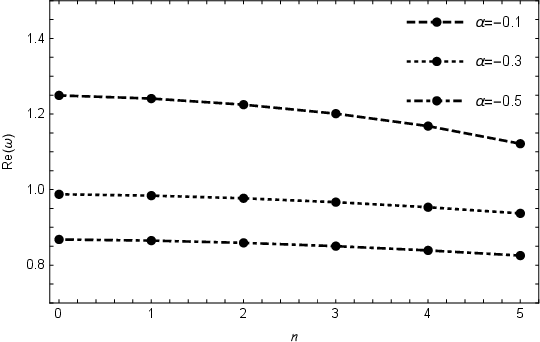}
			\includegraphics[width=0.45\textwidth]{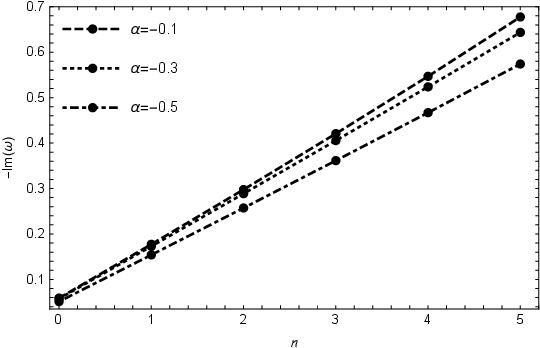}
			\caption {The real part (left panel) and the imaginary part (right panel) of quasinormal frequencies versus $n$ for the massless scalar field with $M=1$, $l=6$ and $g=1$.}
			\label{ReImn}
		\end{center}
	\end{figure*}

\begin{figure*}
		\begin{center}
			\includegraphics[width=0.45\textwidth]{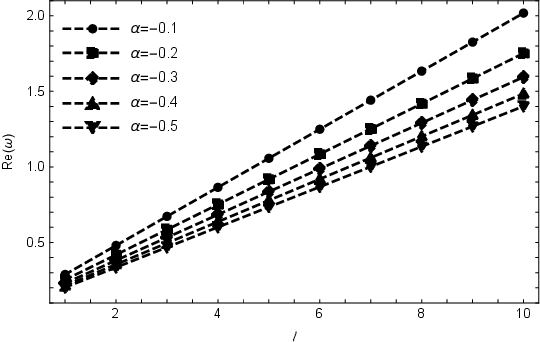}
			\includegraphics[width=0.465\textwidth]{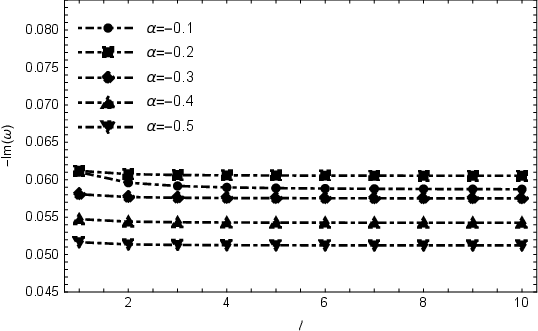}
			\caption {The real part (left panel) and the imaginary part (right panel) of quasinormal frequencies versus $l$ for the massless scalar field with $M=1$, $n=0$ and $g=1$.}
			\label{ReIml}
		\end{center}
	\end{figure*}

\begin{figure*}
		\begin{center}
			\includegraphics[width=0.45\textwidth]{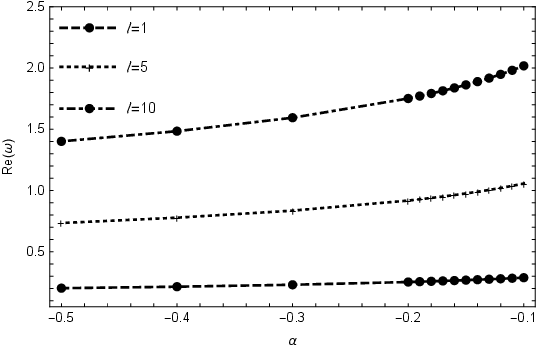}
			\includegraphics[width=0.465\textwidth]{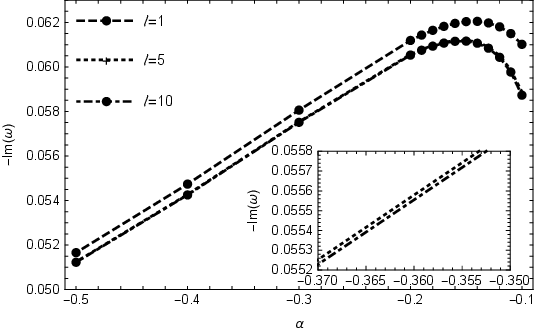}
			\caption {The real part (left panel) and the imaginary part (right panel) of quasinormal frequencies versus $\alpha$ for the massless scalar field with $M=1$, $n=0$ and $g=1$.}
			\label{ReIma}
		\end{center}
	\end{figure*}

\begin{figure*}
		\begin{center}
			\includegraphics[width=0.45\textwidth]{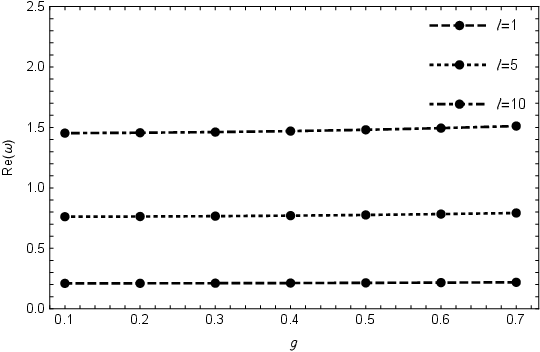}
			\includegraphics[width=0.465\textwidth]{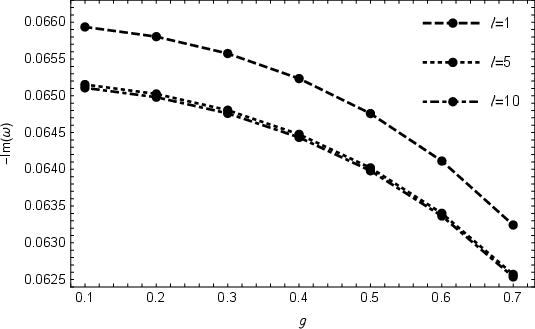}
			\caption {The real part (left panel) and the imaginary part (right panel) of quasinormal frequencies versus $g$ for the massless scalar field with $M=1$, $n=0$ and $\alpha=-0.3$.}
			\label{ReImg}
		\end{center}
	\end{figure*}

	Where $\Lambda_{i}$ denotes a higher-order correction, $\Lambda_2$ and $\Lambda_3$ are given by \cite{Iyer:1986np}
	\begin{eqnarray}
		\Lambda_2 &=& \frac{1}{\sqrt{-2\,V_0''}}\Bigg[\frac{1}{8}
		\left(\frac{V_0^{(4)}}{V_0''}\right)
		\left(\frac{1}{4}+\alpha^2 \right)
		-\frac{1}{288} \left(\frac{V_0^{(3)}}{V_0''}\right)^2
		\left(7+60\alpha^2 \right)\Bigg], \label{ecu.wkbl}\\
		\Lambda_3 &=& \frac{n+\frac{1}{2}}{-2\,V_0''}\Bigg[\frac{5}{6912}
		\left(\frac{V_0^{(3)}}{V_0''}\right)^4 \left(77+188\alpha^2 \right)
		\notag \\
		&&-\frac{1}{384} \left(\frac{V_0'''^2V_0^{(4)}}{V_0''^3}\right)
		\left(51 +100\alpha^2 \right)
		+ \frac{1}{2304}\left(\frac{V_0^{(4)}}{V_0''} \right)^2
		\left(67 +68\alpha^2 \right) \notag\\
		&&+\frac{1}{288} \left(\frac{V_0'''V_0^{(5)}}{V_0''^2} \right)
		\left(19 +28\alpha^2 \right)
		-\frac{1}{288}\left(\frac{V_0^{(6)}}{V_0''}\right)
		\left(5 +4\alpha^2 \right)\Bigg], \label{lamds}
	\end{eqnarray}
	and the other corrections can be found in Ref. ~\cite{Konoplya:2003ii}. In the above equations, $V_0$ represents the maximum of the effective potential and $V^{(n)}_0$ denotes $n$th derivative of $V_0$ to tortoise coordinate $r_*$. Moreover, $\alpha=n+\frac{1}{2}$, and $n$ is overtone number. It is worth pointing out that Konoplya \textit{et al.} explored the importance of overtones for calculating QNMs of regular black holes in asymptotically safe gravity recently\cite{Konoplya:2022hll}. Plugging  Eq.~(\ref{scalarpotential}) into  Eq.~(\ref{w}), we obtain the QNMs of Bardeen black hole surrounded by PFDM for the massless scalar field. Through this process, we can see the changes of the real and imaginary parts of the quasinormal frequencies with different parameters from Figs. \ref{ReImn}-\ref{ReImg}.

    Fig. \ref{ReImn} shows the real part and imaginary part of quasinormal frequencies changing with overtone number $n$ for the massless scalar field when $M=1$, $l=6$, and $g=1$.
	We discover that the real part of the quasinormal frequencies decreases and imaginary part of quasinormal frequencies increases with the increase of $n$ as can be seen in Fig. \ref{ReImn}. It means that the bigger the value of $n$,  the slower the oscillation, and the faster the decay. In other words, $n=0$, i.e., fundamental mode dominates the major contribution to the signal of QNMs \cite{Konoplya:2022hll,Giesler:2019uxc,Oshita:2021iyn}, so we discuss the case of $n=0$ in this paper.
	
	Figure \ref{ReIml} plots the real part and imaginary part of quasinormal frequencies varying with spherical harmonic index $l$ for the massless scalar field when $M=1$, $n=0$, and $g=1$.
	Based on the figure, we obtain that the real part of the quasinormal frequencies monotonously increases while the imaginary part of the quasinormal frequencies decreases with the increase of spherical harmonic index $l$. It means that the increase of spherical harmonic index $l$ causes the massless scalar field to oscillate more quickly and damp more slowly.
	
	Fixed $M=1$, $n=0$, and $g=1$, we depicts the dependence of QNMs on dark matter parameter $\alpha$ for the massless scalar field in Fig. \ref{ReIma}. According to the figure, we find that the increase of dark matter parameter $\alpha$ causes the real part of the quasinormal frequencies increasing, while the imaginary part of the quasinormal frequencies firstly increases and then decreases. It implies that the real oscillation of the quasinormal frequencies increases and the decay rate increases firstly and then decreases as dark matter parameter $\alpha$ increases for the massless scalar field.
	
	Figure \ref{ReImg} draws the behavior of Re $(\omega)$ and -Im $(\omega)$ with respect to magnetic charge $g$ for the massless scalar field when $M=1$, $n=0$, and $\alpha=-0.3$.
	By analyzing the figure, we observe that the magnitude of the real part of the quasinormal frequencies increases and the imaginary part of the quasinormal frequencies decreases as magnetic charge $g$ increases. It indicates that the larger the magnetic charge $g$, the faster the oscillation and the slower the decay for the massless scalar field.

\section{Hawking radiation\label{sec-hawking}}

	In this section, we adopt the WKB method to study the Hawking radiation of Bardeen black hole surrounded by PFDM for the massless scalar field, and discuss the influence of spherical harmonic index, dark matter parameter and magnetic charge on the Hawking radiation.
	
	Taking account of quantum effects, Hawking thought that black holes are not ``black" and can produce thermal radiation, known as Hawking radiation.
	We should note that the difference between radiation at the event horizon which is pure blackbody spectrum and the radiation receiving at infinity which is corrected by the spacetime curvature. The difference can be expressed by a coefficient, i.e., gray-body factor. Using gray-body factor, we can calculate the Hawking radiation of the black hole.
	Therefore, we impose the following boundary conditions on the wave equation:
	\begin{equation}\label{BC}
		\begin{array}{ccll}
			\Psi &=& e^{-i\omega r_*} + R e^{i\omega r_*},& r_* \rightarrow +\infty, \\
			\Psi &=& T e^{-i\omega r_*},& r_* \rightarrow -\infty. \\
		\end{array}%
	\end{equation}
	
	$T$ and $R$ represent transmission and reflection coefficients respectively and their relationship is as follows:
	\begin{equation}\label{tr1}
		|T|^{2}+|R|^{2}=1.
	\end{equation}
	
	Employing the 6th-order WKB method, which describes in Sec.~\ref{sec-massless}, we can get the reflection coefficient
	\begin{equation}\label{r}
		R=\left(1+e^{-2 i \pi K}\right)^{-\frac{1}{2}},
	\end{equation}
	where
	\begin{equation}
		K-i \frac{\left(\omega^{2}-V_{0}\right)}{\sqrt{-2 V_{0}^{\prime \prime}}}-\sum_{i=2}^{i=6} \Lambda_{i}(K)=0.
	\end{equation}

	Then using Eqs.~(\ref{tr1}) and~(\ref{r}), we can obtain the transmission coefficient for each multipole number $l$ easily,
	\begin{equation}
		\left|A_{l}\right|^{2}
		=1-\left|R_{l}\right|^{2}=    \left|T_{l}\right|^{2}.
	\end{equation}
	
	Now, we can calculate the power emission rate of Bardeen black hole surrounded by PFDM for the massless scalar field. We suppose that the black hole and its environment are a canonical ensemble, that is, the Hawking temperature of the black hole does not change between the subsequent emission of two particles \cite{Kanti:2004nr}. Then the power emission rate of the massless scalar field can be written as \cite{Hawking:1975vcx}:
	\begin{equation}
\frac{d^{2} E}{d \omega d t}=\frac{1}{2 \pi} \sum_{l} \frac{N_{l}\left|\mathcal{A}_{l}\right|^{2} \omega}{e^{\omega / T_{H}}-1},
	\end{equation}
	where\cite{Ama-Tul-Mughani:2021tjm}
	\begin{align}\label{wendu}
		T_{H}=\frac{-3 \alpha g^{2} \log \left(\frac{r_{+}}{|\alpha|}\right)+\alpha\left(g^{2}+r_{+}^{2}\right)+r_{+}\left(r_{+}^{2}-2 g^{2}\right)}{4 \pi r_{+}^{2}\left(g^{2}+r_{+}^{2}\right)}.
	\end{align}
	
\begin{figure}
		\begin{minipage}[t]{0.48\linewidth}
			\includegraphics[width=1\linewidth]{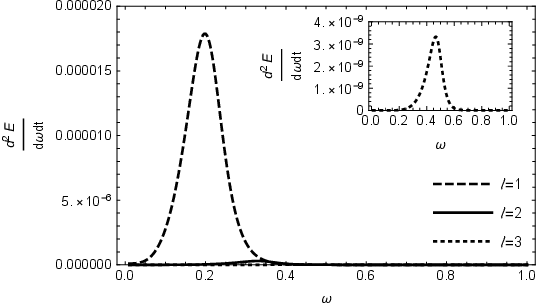}
			\caption{Energy emission rate taken $l$ as the variable with $g=0.1$, $\alpha=-0.3$.}
			\label{fig:h3}
		\end{minipage}
		\hfill
		\begin{minipage}[t]{0.48\linewidth}
			\includegraphics[width=\linewidth]{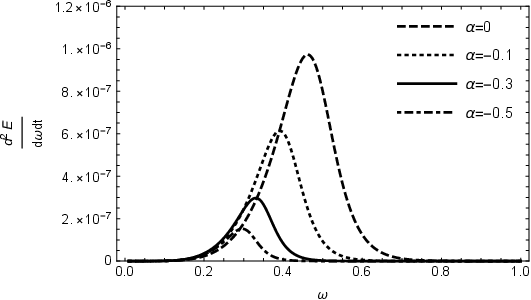}
			\caption{Energy emission rate taken $\alpha$ as the variable with $l=2$, $g=0.1$.}
			\label{fig:h1}
		\end{minipage}
	\end{figure}

	There are $N_{l}=2l+1$ for the massless scalar field and $A_{l}$ representing gray-body factor. $T_{H}$ and $r_{+}$ represent Hawking temperature and event horizon, respectively. Figs. \ref{fig:h3}-\ref{fig:h2} show the energy emission rate with different parameters of Bardeen black hole surrounded by PFDM for the massless scalar field.
	Fixed $g$ and $\alpha$, it is obvious from Fig. \ref{fig:h3}, the energy emission rate is suppressed with the increase of spherical harmonic index $l$ and almost negligible when $l\ge 3$.  Besides, the peak of the energy emission rate shifts to high frequency.
	Then in Fig. \ref{fig:h1}, we consider that the emission of scalar radiation changes with dark matter parameter $\alpha$ when we fix $l$ and $g$. We obtain that the increase of parameter $\alpha$ promotes the emission of scalar radiation and the position of peak moves to right.
	As can be seen from Fig. \ref{fig:h2}, which depicts the effect of magnetic charge $g$ on the power emission spectrum for given $l$ and $\alpha$, we find that the power emission spectrum is depressed as parameter $g$ increases and the location of peak slowly shifts to high frequency.
	We draw a conclusion that the increase of spherical harmonic index $l$ and magnetic charge $g$ both decrease the energy emission rate but dark matter parameter $\alpha$ enhances the energy emission rate. Hence, it is easily to understand that the values of $\alpha$ is smaller and $g$ is larger, the black hole lifespan will be longer.

\section{Absorption cross-section} \label{sec-absorption}
	\begin{figure}
		\begin{minipage}[t]{0.48\linewidth}
			\includegraphics[width=\linewidth]{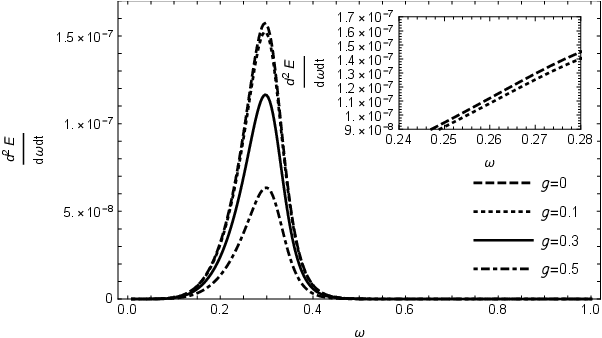}
			\caption{Energy emission rate taken $g$ as the variable with $l=2$, $\alpha=-0.5$.}
			\label{fig:h2}
		\end{minipage}
		\hfill%
		\begin{minipage}[t]{0.48\linewidth}
			\includegraphics[width=0.875\linewidth]{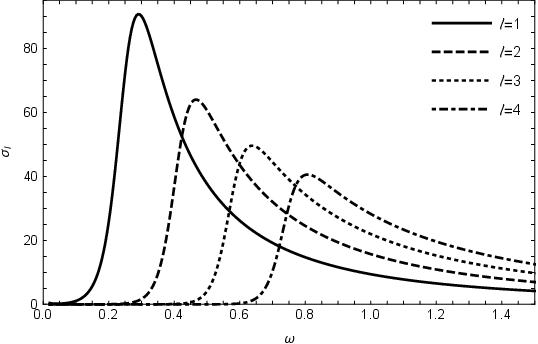}
			\caption{The partial absorption cross-section taken $l$ as the variable with $\alpha=-0.1$, $g=0.1$.}
			\label{fig:abl1}
		\end{minipage}
	\end{figure}	
	
	In 2014, Benone \textit{et al.} \cite{Benone:2014qaa} developed the partial wave method to compute the absorption cross-section, which uses the transmission coefficient $T_{l}$ in process of calculation.
	In this section, we adopt the partial wave method to calculate the absorption cross-section of Bardeen black hole surrounded by PFDM for the massless scalar field. The total absorption cross-section can be written as:
	\begin{equation}
		\sigma_{total}=\sum_{l=0}^{\infty} \sigma_{l},
	\end{equation}
	and the expression of the partial absorption cross-section is
	\begin{equation}
		\sigma_{l} = \frac{\pi (2l+1)}{\omega^2} \left(1-\left|\mathrm{e}^{2 \mathrm{i} \delta_{l}}\right|^{2}\right)
		=\frac{\pi (2l+1)}{\omega^2} \left|T_{l}\right|^{2} .
	\end{equation}
	\begin{figure}
		\begin{minipage}[t]{0.45\linewidth}
			\includegraphics[width=\linewidth]{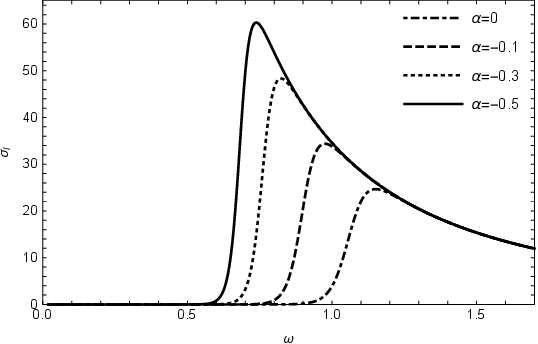}
			\caption{The partial absorption cross-section taken $\alpha$ as the variable with $g=0.1$, $l=5$.}
			\label{fig:bpb1}
		\end{minipage}
		\hfill
		\begin{minipage}[t]{0.45\linewidth}
			\includegraphics[width=\linewidth]{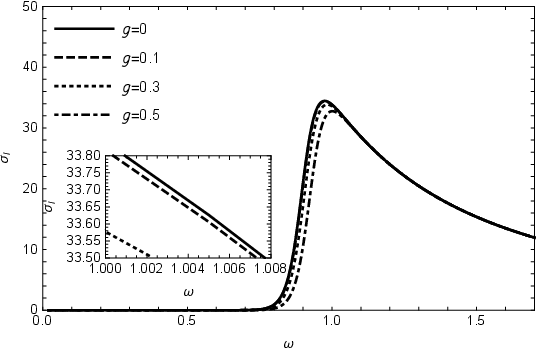}
			\caption{The partial absorption cross-section taken $g$ as the variable with $\alpha=-0.1$, $l=5$.}
			\label{fig:bpschp1}
		\end{minipage}
	\end{figure}	
	In Fig. \ref{fig:abl1}, we exhibit the partial absorption cross-section of Bardeen black hole surrounded by PFDM for the massless scalar field with fixed $\alpha=-0.1$ and $g=0.1$. Analyzing the graph, we discover that with the increase of frequency, the partial absorption cross-section first starts from zero and climbs the peak, and then diminishes. Besides, the maximum value of the partial absorption cross-section decreases as $l$ increases.
	For exploring the influence of dark matter parameter $\alpha$ on the partial absorption cross-section, we show the behavior of the partial absorption cross-section with different values of $\alpha$ and take $g=0.1$ and $l=5$ in Fig. \ref{fig:bpb1}. As shown in the figure, the bigger the value of $\alpha$, the lower the partial absorption cross-section. It means that dark matter parameter suppresses the partial absorption cross-section. When $\alpha=-0.1$ and $l=5$, the curves of the partial absorption cross-section for the different values of magnetic charge $g$ are presented in Fig. \ref{fig:bpschp1}. It can be seen that the peak value of the partial absorption cross-section decreases with the increase of $g$. In other words, magnetic charge also depresses the partial absorption cross-section.
	On the basis of the above discussion, we obtain that $l$, $\alpha$ and $g$ all suppress the partial absorption cross-section. The result is compatible to the variation of effective potential, which is higher with $l$, $\alpha$ and $g$ increasing. And the essence of the phenomenon is that the higher potential barrier causes the less massless scalar wave from infinity transmitting to black hole.

\begin{figure}
		\begin{minipage}[t]{0.45\linewidth}
			\includegraphics[width=\linewidth]{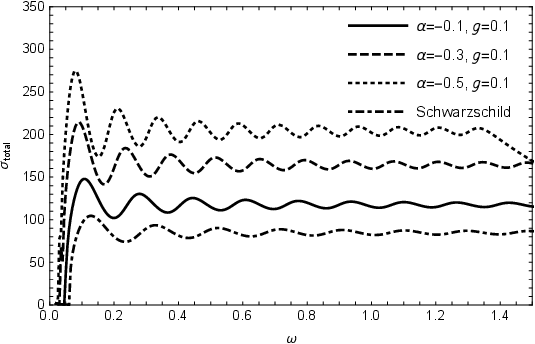}
			\caption{The total absorption cross-section taken $\alpha$ as the variable with $g=0.1$.}
			\label{fig:aba}
		\end{minipage}
		\hfill
		\begin{minipage}[t]{0.45\linewidth}
			\includegraphics[width=\linewidth]{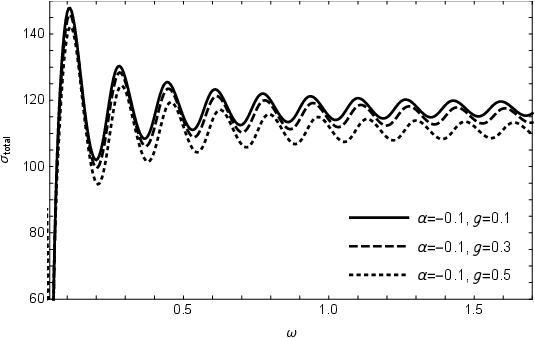}
			\caption{The total absorption cross-section taken $g$ as the variable with $\alpha=-0.1$.}
			\label{fig:abg}
		\end{minipage}
	\end{figure}

	To better investigate the properties of the considered black hole, we also depict the total absorption cross-section for the massless scalar field in Figs. \ref{fig:aba} and \ref{fig:abg}, which contributes from $l=0$ up to $l$=10. It can be observed that the total absorption cross-section decreases as $\alpha$ and $g$ increase. Its variation is consistent with the partial absorption cross-section.
	
\begin{figure}
		\begin{minipage}[t]{0.45\linewidth}
			\includegraphics[width=\linewidth]{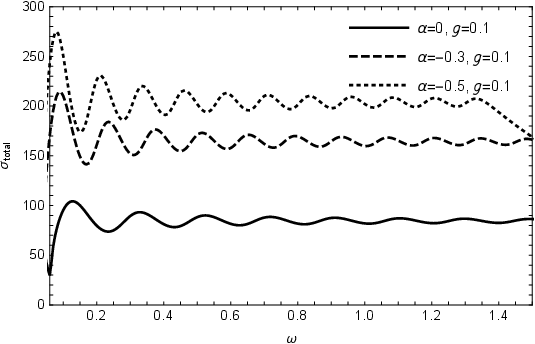}
			\caption{The total absorption cross-section of Bardeen black hole surrounded by PFDM ($\alpha\neq0$) and Bardeen black hole ($\alpha=0$) with $g=0.1$.}
			\label{fig:bpb}
		\end{minipage}
		\hfill
		\begin{minipage}[t]{0.45\linewidth}
			\includegraphics[width=\linewidth]{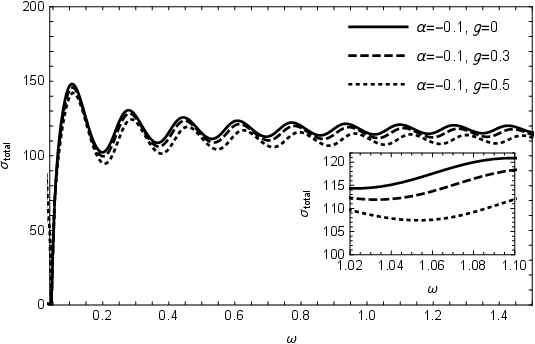}
			\caption{The total absorption cross-section of Bardeen black hole surrounded by PFDM ($g\neq0$) and Schwarzschild black hole surrounded by PFDM ($g=0$) with $\alpha=-0.1$.}
			\label{fig:bpscp}
		\end{minipage}
	\end{figure}

	Fig. \ref{fig:bpb} is drawn to compare the total absorption cross-section of the Bardeen black hole surrounded by PFDM ($\alpha\neq0$) with Bardeen black hole ($\alpha=0$). We find that the total absorption cross-section increases because of the presence of $\alpha$. In Fig. \ref{fig:bpscp}, we compare the total absorption cross-section of Bardeen black hole surrounded by PFDM ($g\neq0$) with Schwarzschild black hole surrounded by PFDM ($g=0$), and the figure shows that the total absorption cross-section diminishes due to the contribution of $g$. By comparing Figs. \ref{fig:bpb} and \ref{fig:bpscp}, it is found that the influence of dark matter parameter on the total absorption cross-section is greater than that of the magnetic charge.

	\section{Conclusion}\label{sec-conclusion}
	
	In summary, we have investigated QNMs, Hawking radiation and absorption cross-section for the massless scalar field of Bardeen black hole surrounded by PFDM, which is described by dark matter parameter and magnetic charge. Firstly, we have discussed the effective potential taking $l$, $\alpha$, and $g$ as variables. Our analysis implies that the barrier height of effective potential increases with the increase of $l$, $\alpha$ and $g$.
	Then, at the situation of fundamental mode, we have carefully studied QNMs of the considered black hole for the massless scalar field with the help of 6th-order WKB method.
	It has been discovered that the real part of quasinormal frequencies becomes more positive whereas the imaginary part decreases when the value of spherical harmonic index $l$ increases. This indicates that the actual frequency of the oscillation increases and the damping decreases with $l$ increasing.
	And it has been found that the real part of quasinormal frequencies increases with dark matter parameter $\alpha$ increasing but the imaginary part first increases and then decreases. In other words, the oscillation frequency becomes faster and the decay rate shows a trend of first increasing then decreasing when the value of $\alpha$ increases.
	As magnetic charge $g$ increases, the real part of quasinormal frequencies increases while the imaginary part reduces. Namely, $g$ also enhances the actual frequency of the oscillation and depresses decay rate.
	
	Moreover, we have calculated the power emission rate of the black hole for the massless scalar field using 6th-order WKB method. The result shows that the energy emission rate is impeded by the increase of multiple number $l$ and magnetic charge $g$ while dark matter parameter $\alpha$ enhances it.
	And the location of peak shifts to high frequency as the three parameters increase separately.

	Finally, we have devoted to computing absorption cross-section for the massless scalar field by employing the partial wave method. It is obvious that $l$, $\alpha$, and $g$ depress the partial cross-section which corresponds to the dependence of the effective potential increases with $l$, $\alpha$ and $g$ increasing, respectively. It implies that the higher the height of effective potential, the less waves can transmit. In further analysis, we have compared the total absorption cross-section of three black holes, i.e., Bardeen black hole, Schwarzschild black hole surrounded by PFDM and Bardeen black hole surrounded by PFDM from $l=0$ to $l=10$. Fixed $g=0.1$, we have explored the existence of dark matter parameter affecting the total absorption cross-section. It has been investigated that the total absorption cross-section of Bardeen black hole surrounded by PFDM ($\alpha\neq0$) is bigger than that of Bardeen black hole ($\alpha=0$). And when $\alpha=-0.1$, we have studied the existence of magnetic charge influencing the total absorption cross-section. It has been found that the total absorption cross-section of Bardeen black hole surrounded by PFDM ($g\neq0$) is smaller than that of Schwarzschild black hole surrounded by PFDM ($g=0$). That is to say, the presence of $\alpha$ enhances the total absorption cross-section and the presence of $g$ decreases it.

	\section*{Acknowledgments}
	This work was supported by the National Natural Science Foundation of China (Grant No. 12065012), Yunnan Fundamental Research Projects (Grant No. 202301AS070029), and Yunnan High-level Talent Training Support Plan Young \& Elite Talents Project (Grant No. YNWR-QNBJ-2018-360).


\begin{thebibliography}{}
		\bibitem{LIGOScientific:2016aoc}
		B.~P.~Abbott \textit{et al.},
		{\it Phys. Rev. Lett.} \textbf{116}, 061102 (2016).
		
		
		\bibitem{LIGOScientific:2016sjg}
		B.~P.~Abbott \textit{et al.},
		{\it Phys. Rev. Lett.} \textbf{116}, 241103 (2016).
		
		
		
		
		\bibitem{LIGOScientific:2018mvr}
		B.~P.~Abbott \textit{et al.},
		{\it Phys. Rev. X} \textbf{9}, 031040 (2019).
		
		
		
		
		\bibitem{EventHorizonTelescope:2019dse}
		K.~Akiyama \textit{et al.},
		{\it Astrophys. J. Lett.} \textbf{875}, L1 (2019).
		
		
		\bibitem{EventHorizonTelescope:2019ths}
		K.~Akiyama \textit{et al.},
		{\it Astrophys. J. Lett.} \textbf{875}, L4 (2019).
		
		\bibitem{Bardeen}
		J. M. Bardeen,
		{\it In Proc. Int. Conf. GR5, Tbilisi}, 174 (1968).
		
		
		
		
		
		
		\bibitem{Ayon-Beato:2000mjt}
		E.~Ayon-Beato and A.~Garcia,
		{\it Phys. Lett. B} \textbf{493}, 149 (2000).
		
		
		\bibitem{Dymnikova:1992ux}
		I.~Dymnikova,
		{\it Gen. Relativ. Gravit.} \textbf{24}, 235 (1992).
		
		
		\bibitem{Bronnikov:2000vy}
		K.~A.~Bronnikov,
		{\it Phys. Rev. D} \textbf{63}, 044005 (2001).
		
		
		
		\bibitem{Hayward:2005gi}
		S.~A.~Hayward,
		{\it Phys. Rev. Lett.} \textbf{96}, 031103 (2006).
		

        \bibitem{Fan:2016hvf}
        Z.~Y.~Fan and X.~Wang,
        {\it Phys. Rev. D} \textbf{94}, 124027 (2016).


\bibitem{Toshmatov:2017zpr}
B.~Toshmatov, Z.~Stuchl\'\i{}k and B.~Ahmedov,
{\it Phys. Rev. D} \textbf{95}, 084037 (2017).








		
		
		\bibitem{Tsukamoto:2017fxq}
		N.~Tsukamoto,
		{\it Phys. Rev. D} \textbf{97}, 064021 (2018).
		
		\bibitem{Kumar:2019pjp}
		R.~Kumar, S.~G.~Ghosh and A.~Wang,
		{\it Phys. Rev. D} \textbf{100}, 124024 (2019).
		
		
		\bibitem{Ahmed:2020jic}
		F.~Ahmed, D.~V.~Singh and S.~G.~Ghosh,
		{\it Gen. Relativ. Gravit.} \textbf{54}, 21 (2022).
		
		
		\bibitem{Saadat:2013rna}
		H.~Saadat,
		{\it Int. J. Theor. Phys.} \textbf{52}, 3255 (2013).
		
		
		\bibitem{Breton:2015iev}
		N.~Bret\'on and S.~E.~Perez Bergliaffa,
		{\it AIP Conf. Proc.} \textbf{1577}, 112 (2015).
		
		
		
		\bibitem{Jawad:2020hju}
		A.~Jawad, M.~Yasir and S.~Rani,
		{\it Mod. Phys. Lett. A} \textbf{35}, 2050298 (2020).
		
		
		
		
		
		
		\bibitem{OortJH}
		J.~H.~Oort,
		{\it Bull. Astron. Inst. Neth.} \textbf{6}, 249 (1932).
		
		
		
		\bibitem{Rubin:1970zza}
		V.~C.~Rubin and W.~K.~Ford, Jr.,
		{\it Astrophys. J.} \textbf{159}, 379 (1970).
		
		
		\bibitem{Clowe:2006eq}
		D.~Clowe, M.~Bradac, A.~H.~Gonzalez, M.~Markevitch, S.~W.~Randall, C.~Jones and D.~Zaritsky,
		{\it Astrophys. J. Lett.} \textbf{648}, L109 (2006).
		
		
		
		\bibitem{Planck:2015fie}
		P.~A.~R.~Ade \textit{et al.} [Planck],
		{\it Astron. Astrophys.} \textbf{594}, A13 (2016).
		
		
		\bibitem{Jamil:2009eb}
		M.~Jamil, E.~N.~Saridakis and M.~R.~Setare,
		{\it Phys. Rev. D} \textbf{81}, 023007 (2010).
		
		
		\bibitem{Pantig:2022toh}
		R.~C.~Pantig and A.~\"Ovg\"un,
		{\it Eur. Phys. J. C} \textbf{82}, 391 (2022).
		
		
		
		
		\bibitem{Pantig:2021zqe}
		R.~C.~Pantig, P.~K.~Yu, E.~T.~Rodulfo and A.~\"Ovg\"un,
		{\it Ann Phys (N Y).} \textbf{436}, 168722 (2022).




\bibitem{Konoplya:2019sns}
R.~A.~Konoplya,
{\it Phys. Lett. B} \textbf{795}, 1 (2019).



\bibitem{Stuchlik:2022xtq}
Z.~Stuchl\'\i{}k and J.~Vrba,
{\it Astrophys. J.} \textbf{935}, 91 (2022).



		
		
		
		\bibitem{Kiselev:2002dx}
		V.~V.~Kiselev,
		{\it Class. Quant. Grav.} \textbf{20}, 1187 (2003).

\bibitem{Kiselev:2003ah}
V.~V.~Kiselev,
[arXiv:gr-qc/0303031 [gr-qc]].



		
\bibitem{Toshmatov:2015npp}
B.~Toshmatov, Z.~Stuchl\'\i{}k and B.~Ahmedov,
{\it Eur. Phys. J. Plus} \textbf{132}, 98 (2017).

\bibitem{Chakrabarty:2018skk}
H.~Chakrabarty, A.~A.~Abdujabbarov and C.~Bambi,
{\it Eur. Phys. J. C} \textbf{79}, 179 (2019).





		
		\bibitem{Heydarzade:2017wxu}
		Y.~Heydarzade and F.~Darabi,
		{\it Phys. Lett. B} \textbf{771}, 365 (2017).
		
		
		\bibitem{Xu:2017vse}
		Z.~Xu, X.~Hou, X.~Gong and J.~Wang,
		{\it Eur. Phys. J. C} \textbf{78}, 513 (2018).
		
		
		\bibitem{Zhang:2020mxi}
		H.~X.~Zhang, Y.~Chen, T.~C.~Ma, P.~Z.~He and J.~B.~Deng,
		{\it Chin. Phys. C} \textbf{45}, 055103 (2021).
		
		
		
		
		
		\bibitem{Berti:2005ys}
		E.~Berti, V.~Cardoso and C.~M.~Will,
		{\it Phys. Rev. D} \textbf{73}, 064030 (2006).
		
		
		
		\bibitem{Vishveshwara:1970zz}
		C.~V.~Vishveshwara,
		{\it Nature} \textbf{227}, 936 (1970).
		
		
		\bibitem{Chandrasekhar:1975zza}
		S.~Chandrasekhar and S.~L.~Detweiler,
		{\it Proc. Roy. Soc. Lond. A} \textbf{344}, 441 (1975).
		
		
		
		\bibitem{Setare:2003hm}
		M.~R.~Setare,
		{\it Class. Quant. Grav.} \textbf{21}, 1453 (2004).
		
		
		
		\bibitem{Setare:2003bd}
		M.~R.~Setare,
		{\it Phys. Rev. D} \textbf{69}, 044016 (2004).
		
		
		
		
		\bibitem{Setare:2004rt}
		M.~R.~Setare,
		{\it Gen. Relativ. Gravit.} \textbf{37}, 1411 (2005).
		
		
		
		\bibitem{Setare:2004uu}
		M.~R.~Setare and E.~C.~Vagenas,
		{\it Mod. Phys. Lett. A} \textbf{20}, 1923 (2005).
		
		
		\bibitem{Zhang:2006ij}
		Y.~Zhang and Y.~X.~Gui,
		{\it Class. Quant. Grav.} \textbf{23}, 6141 (2006).
		
		
		\bibitem{Stefanov:2010xz}
		I.~Z.~Stefanov, S.~S.~Yazadjiev and G.~G.~Gyulchev,
		{\it Phys. Rev. Lett.} \textbf{104}, 251103 (2010).
		
		
		\bibitem{Stefanov:2010qwy}
		I.~Z.~Stefanov, S.~S.~Yazadjiev and G.~N.~Gyulchev,
		{\it AIP Conf. Proc.} \textbf{1301}, 355 (2010).

\bibitem{Toshmatov:2015wga}
B.~Toshmatov, A.~Abdujabbarov, Z.~Stuchl\'\i{}k and B.~Ahmedov,
{\it Phys. Rev. D} \textbf{91}, 083008 (2015).

		
		
		\bibitem{Konoplya:2020bxa}
		R.~A.~Konoplya and A.~F.~Zinhailo,
		{\it Eur. Phys. J. C} \textbf{80}, 1049 (2020).
		
		
		\bibitem{Ma:2022gzr}
		C.~Ma, Y.~Zhang, Q.~Li and Z.~W.~Lin,
		{\it Commun. Theor. Phys.} \textbf{74}, 065402 (2022).
		
		
		
		\bibitem{Konoplya:2020jgt}
		R.~A.~Konoplya, A.~F.~Zinhailo and Z.~Stuchlik,
		{\it Phys. Rev. D} \textbf{102}, 044023 (2020).
		
		
		
		
		\bibitem{Pantig:2022gih}
		R.~C.~Pantig, L.~Mastrototaro, G.~Lambiase and A.~\"Ovg\"un,
		{\it Eur. Phys. J. C} \textbf{82}, 1155 (2022).
		
		
		
		
		
		
		
		
		
		
		
		
		
		\bibitem{Hawking:1975vcx}
		S.~W.~Hawking,
		{\it Commun. Math. Phys.} \textbf{43}, 199 (1975)
		[erratum: Commun. Math. Phys. \textbf{46}, 206 (1976)].
		
		\bibitem{Damour:1976jd}
		T.~Damour and R.~Ruffini,
		{\it Phys. Rev. D} \textbf{14}, 332 (1976).
		
		
		\bibitem{Parikh:1999mf}
		M.~K.~Parikh and F.~Wilczek,
		{\it Phys. Rev. Lett.} \textbf{85}, 5042 (2000).
		
		
		\bibitem{Kerner:2007rr}
		R.~Kerner and R.~B.~Mann,
		{\it Class. Quant. Grav.} \textbf{25}, 095014 (2008).
		
		
		
		\bibitem{Kerner:2008qv}
		R.~Kerner and R.~B.~Mann,
		{\it Phys. Lett. B} \textbf{665}, 277 (2008).
		
		
		
		
		\bibitem{Konoplya:2003ii}
		R.~A.~Konoplya,
		{\it Phys. Rev. D} \textbf{68}, 024018 (2003).
		
		
		\bibitem{Kokkotas:2010zd}
		K.~D.~Kokkotas, R.~A.~Konoplya and A.~Zhidenko,
		{\it Phys. Rev. D} \textbf{83}, 024031 (2011).
		
		\bibitem{Konoplya:2020cbv}
		R.~A.~Konoplya and A.~F.~Zinhailo,
		{\it Phys. Lett. B} \textbf{810}, 135793 (2020).
		
		
		
		\bibitem{Li:2022wzi}
		Q.~Li, C.~Ma, Y.~Zhang, Z.~W.~Lin and P.~F.~Duan,
		{\it Eur. Phys. J. C} \textbf{82}, 658 (2022).
		
		
		
		
		
		\bibitem{Li:2022jda}
		Q.~Li, C.~Ma, Y.~Zhang, Z.~W.~Lin and P.~F.~Duan,
		{\it Chin. J. Phys.} \textbf{77}, 1269 (2022).
		
		
		
		\bibitem{Okyay:2021nnh}
		M.~Okyay and A.~\"Ovg\"un,
		{\it JCAP} \textbf{1}, 009 (2022).
		
		
		\bibitem{Pantig:2022ely}
		R.~C.~Pantig and A.~\"Ovg\"un,
		{\it Ann Phys (N Y).} \textbf{448}, 169197 (2023).
		
		
		
		\bibitem{Sanchez:1977si}
		N.~G.~Sanchez,
		{\it Phys. Rev. D} \textbf{18}, 1030 (1978).
		
		
		\bibitem{Unruh:1976fm}
		W.~G.~Unruh,
		{\it Phys. Rev. D} \textbf{14}, 3251 (1976).
		
		
		\bibitem{Crispino:2007qw}
		L.~C.~B.~Crispino, E.~S.~Oliveira, A.~Higuchi and G.~E.~A.~Matsas,
		{\it Phys. Rev. D} \textbf{75}, 104012 (2007).
		
		
		
		\bibitem{Macedo:2014uga}
		C.~F.~B.~Macedo and L.~C.~B.~Crispino,
		{\it Phys. Rev. D} \textbf{90}, 064001 (2014).
		
		
		
		
		
		\bibitem{Li:2012zx}
		M.~H.~Li and K.~C.~Yang,
		{\it Phys. Rev. D} \textbf{86}, 123015 (2012).
		
		
		
		\bibitem{Schutz:1985km}
		B.~F.~Schutz and C.~M.~Will,
		{\it Astrophys. J. Lett.} \textbf{291}, L33 (1985).
		
		\bibitem{Iyer:1986np}
		S.~Iyer and C.~M.~Will,
		{\it Phys. Rev. D} \textbf{35}, 3621 (1987).
		
		
		
		
		
		\bibitem{Matyjasek:2017psv}
		J.~Matyjasek and M.~Opala,
		{\it Phys. Rev. D} \textbf{96}, 024011 (2017).
		
\bibitem{Konoplya:2022hll}
R.~A.~Konoplya, A.~F.~Zinhailo, J.~Kunz, Z.~Stuchlik and A.~Zhidenko,
{\it JCAP} \textbf{10}, 091 (2022).

\bibitem{Giesler:2019uxc}
M.~Giesler, M.~Isi, M.~A.~Scheel and S.~Teukolsky,
{\it Phys. Rev. X} \textbf{9}, 041060 (2019).

\bibitem{Oshita:2021iyn}
N.~Oshita,
{\it Phys. Rev. D} \textbf{104}, 124032 (2021).






		
		\bibitem{Kanti:2004nr}
		P.~Kanti,
		{\it Int. J. Mod. Phys. A} \textbf{19}, 4899  (2004).
		
		
		\bibitem{Ama-Tul-Mughani:2021tjm}
		Q.~Ama-Tul-Mughani, W.~us Salam and R.~Saleem,
		{\it Astropart. Phys.} \textbf{132}, 102623 (2021).
		
		
		
		
		\bibitem{Benone:2014qaa}
		C.~L.~Benone, E.~S.~de Oliveira, S.~R.~Dolan and L.~C.~B.~Crispino,
		{\it Phys. Rev. D} \textbf{89}, 104053 (2014).
\end{thebibliography}
\end{document}